\documentclass[review,onecolumn]{elsarticle}
\usepackage{amsmath}
\usepackage{lineno,hyperref}
\journal{Journal of \LaTeX\ Templates}
\begin{document}

\begin{frontmatter}
\title{Second-harmonic generation in the system with fractional diffraction}
\author{Pengfei Li$^{a,*}$}
\cortext[mycorrespondingauthor]{Corresponding author}
\ead{lipf@tynu.edu.cn}
\author{Hidetsugu Sakaguchi$^{b}$}
\author{Liangwei Zeng$^{c}$}
\author{Xing Zhu$^{c}$}
\author{Dumitru Mihalache$^{d}$}
\author{Boris A. Malomed$^{e,f}$}

\address{$^{a}$Department of Physics, Taiyuan Normal University, Jinzhong, 030619, China}
\address{$^{b}$Interdisciplinary Graduate School of Engineering Sciences, Kyushu University, Kasuga, Fukuoka 816-8580, Japan}
\address{$^{c}$Department of Basic Course, Guangzhou Maritime University, Guangzhou 510725, China}
\address{$^{d}$Horia Hulubei National Institute of Physics and Nuclear Engineering, Magurele, Bucharest, RO-077125, Romania}
\address{$^{e}$Department of Physical Electronics, School of Electrical Engineering, Faculty of Engineering, and Center for Light-Matter Interaction, Tel Aviv University, Tel Aviv 69978, Israel}
\address{$^{f}$Instituto de Alta Investigaci\'{o}n, Universidad de Tarapac\'{a}, Casilla 7D, Arica, Chile}

\begin{abstract}
We construct a family of bright optical solitons composed of fundamental-frequency
(FF) and second-harmonic (SH) components in the one-dimensional (planar) waveguide with
the quadratic (second-harmonic-generating) nonlinearity and effective fractional diffraction,
characterized by the
L\'{e}vy index $\alpha$, taking values between $2$ and $0.5$, which correspond to the
non-fractional diffraction and critical collapse, respectively. The existence domain and
stability boundary for the solitons are delineated in the space of $\alpha$, FF-SH
mismatch parameter, and propagation constant. The stability boundary is tantamount to
that predicted by the Vakhitov-Kolokolov criterion, while unstable solitons spontaneously
evolve into localized breathers. A sufficiently weak transverse kick applied to the stable
solitons excite small internal vibrations in the stable solitons, without setting them in motion.
A stronger kick makes the solitons' trajectories tilted, simultaneously destabilizing the solitons.


\end{abstract}

\end{frontmatter}

\section{Introduction}

\label{Sec I}

The concept of the fractional spatial dispersion has been first brought to
the forefront of research in physics in the framework of fractional quantum
mechanics for particles moving by L\'{e}vy flights \cite{Laskin1,Laskin2,Laskin3,Book-Laskin}.
Particular implementations of the
fractional Schr\"{o}dinger equations were proposed in L\'{e}vy crystals \cite{Levy-Crystal} and
polariton condensates \cite{Polariton-conden}, although
experimental realization of the fractional quantum mechanics has not been
reported, as yet. On the other hand, it was proposed to make use of the
commonly known similarity of the Schr\"{o}dinger equation and the classical
equation for the paraxial light propagation, and implement the fractional
diffraction in optics, which will emulate the action of the fractional
kinetic-energy operator in the quantum theory \cite{FD-Optics1}. The
proposal relied on using the \textit{4f} optical setup to split the light
beam into spectral components corresponding to different values of the
transverse wave number by means of a lens, passing the separated components
through an appropriately designed phase plate, which should impart to them
phase shifts emulating the action of the fractional diffraction, and
recombining the components back into a single beam by means of the second
lens.

In the optical system, it is natural to combine the fractional diffraction
with the nonlinearity of optical materials, which leads to the introduction
of various forms of the fractional nonlinear Schr\"{o}dinger (NLS) equations
and prediction of a variety of fractional solitons and nonlinear phenomena,
such as accessible \cite{fNLSE1,fNLSE4}, gap \cite{fNLSE2,fNLSE7,fNLSE8}, and
surface \cite{fNLSE5,fNLSE11} solitons, solitary vortices \cite{fNLSE9}, fractional solitons in
nonlinear lattice \cite{fNLSE10}, wave
collapse \cite{fNLSE3}, and other nonlinear effects \cite{fNLSE6}-\cite{fNLSE20}.
Another natural extension was developed for optical waveguides
and cavities combining fractional diffraction with losses, gain, and
nonlinearity, leading to models in the form of fractional NLS equations with
$\mathcal{PT}$-symmetric complex potentials, that also give rise to a
variety of soliton states \cite{fNLSE-PT1}-\cite{fNLSE-PT8}. Further,
solitons \cite{CFNLSE1}-\cite{CFNLSE3} and domain-wall states \cite{DW-FNLSE}
were predicted as solutions of systems of coupled fractional NLS equations.
A part of these findings were summarized in Ref. \cite{Malomed-Rev}.

An important branch of nonlinear optics, including the formation of
solitons, is based on the second-harmonic generation in media with quadratic
($\chi ^{\left( 2\right) }$) nonlinearity \cite{Torruellas}-\cite{SH6}. In
this work, we systematically investigate the formation and stability of
bright solitons in the system with the fractional diffraction acting on both
the fundamental-frequency (FF) and second-harmonic (SH) waves coupled by the
$\chi ^{\left( 2\right) }$ interaction.

The rest of the paper is organized as follows. The model is introduced in
Section \ref{Sec II}. It includes the considerations of the cascading limit,
which replaces the quadratic nonlinearity by an effective cubic one.
Systematic results for families of stationary solitons and their stability
are reported in Section \ref{Sec III}. Dynamics of the solitons is addressed
in Section \ref{Sec IV}. The paper is concluded by Section \ref{Sec V}.

\section{The model}

\label{Sec II}

\subsection{The fractional system with the $\protect\chi ^{(2)}$ interaction}

The starting point is the system of one-dimensional coupled nonlinear
equations for amplitudes of the FF and SH waves, $\Psi _{1}\left( x,z\right)
$ and $\Psi _{2}\left( x,z\right) $:
\begin{equation}
i\frac{\partial \Psi _{1}}{\partial z}-D_{1}\left( -\frac{\partial ^{2}}{%
\partial x^{2}}\right) ^{\alpha /2}\Psi _{1}+\Psi _{1}^{\ast }\Psi _{2}=0,
\label{system_Psi1}
\end{equation}%
\begin{equation}
2i\frac{\partial \Psi _{2}}{\partial z}-D_{2}\left( -\frac{\partial ^{2}}{%
\partial x^{2}}\right) ^{\alpha /2}\Psi _{2}+Q\Psi _{2}+\frac{1}{2}\Psi
_{1}^{2}=0,  \label{system_Psi2}
\end{equation}%
where $z$ is the propagation distance, $x$ is the transverse coordinate,
while $D_{1}$ and $D_{2}$ are the FF and SH diffraction coefficients. Real $%
Q $ is the mismatch parameter of the $\chi ^{\left( 2\right) }$\
interaction, with $\ast $ standing for the complex conjugate. The fractional
diffraction is represented by Riesz derivative with L\'{e}vy index (LI) $%
\alpha $ \cite{Riesz,Riesz2},%
\begin{equation}
\left( -\frac{\partial ^{2}}{\partial x^{2}}\right) ^{\alpha /2}\Psi (x)=%
\frac{1}{2\pi }\int_{-\infty }^{+\infty }|p|^{\alpha }dp\int_{-\infty
}^{+\infty }d\xi e^{ip(x-\xi )}\Psi (\xi ).  \label{Risz derivative}
\end{equation}%
Normally, LI takes values $1<\alpha \leq 2$, but it is also possible to
consider values $0<\alpha \leq 1$. The usual diffraction corresponds to $%
\alpha =2$. Straightforward analysis demonstrates that Eqs. (\ref%
{system_Psi1}) and (\ref{system_Psi2}) with the quadratic nonlinearity do
not produce the collapse in the interval of%
\begin{equation}
1/2<\alpha \leq 2.  \label{no-coll}
\end{equation}%
The critical and supercritical collapse occurs at $\alpha =1/2$ and $\alpha
<1/2$, respectively (recall that the combination of the fractional
diffraction with the cubic nonlinearity gives rise to the collapse at $%
\alpha \leq 1$ \cite{Malomed-Rev}).

Stationary solutions to Eqs. (\ref{system_Psi1}) and (\ref{system_Psi2})
with FF and SH propagation constants $\beta _{1}$ and $\beta _{2}\equiv
2\beta _{1}$ are looked for as%
\begin{equation}
\Psi _{1}\left( x,z\right) =e^{i\beta _{1}z}\psi _{1}(x),\Psi _{2}\left(
x,z\right) =e^{2i\beta _{1}z}\psi _{2}(x),  \label{Psi1Psi2}
\end{equation}%
where the real functions $\psi _{1,2}(x)$ satisfy the following system of
stationary equations:
\begin{equation}
-D_{1}\left( -\frac{\partial ^{2}}{\partial x^{2}}\right) ^{\alpha /2}\psi
_{1}-\beta _{1}\psi _{1}+\psi _{1}^{\ast }\psi _{2}=0,  \label{system_psi1}
\end{equation}

\begin{equation}
-D_{2}\left( -\frac{\partial ^{2}}{\partial x^{2}}\right) ^{\alpha /2}\psi
_{2}+Q\psi _{2}-4\beta _{1}\psi _{2}+\frac{1}{2}\psi _{1}^{2}=0.
\label{system_psi2}
\end{equation}%
Two-wave soliton families may be naturally characterized, as usual, by
dependences between the propagation constant $\beta _{1}$ and the power
(alias the\ Manley-Rowe invariant, or the soliton's norm),
\begin{equation}
N=\int_{-\infty }^{+\infty }\left( \left\vert \psi _{1}\right\vert
^{2}+4\left\vert \psi _{2}\right\vert ^{2}\right) dx,  \label{N}
\end{equation}%
which is a dynamical invariant of the underlying system of Eqs. (\ref%
{system_Psi1}) and (\ref{system_Psi2}).

\subsection{The cascading limit for $Q<0$ and the scaling invariance for $%
Q=0 $}

The approximation known as the cascading limit (CL) corresponds, roughly
speaking, to large values of $\left\vert Q\right\vert $. In this case, one
neglects the fractional-derivative and propagation-constant terms in
comparison with other linear terms in Eq. (\ref{system_psi2}), which yields
\
\begin{equation}
\psi _{2}\approx -\left( 2Q\right) ^{-1}\psi _{1}^{2}.  \label{CL1}
\end{equation}%
The substitution of approximation (\ref{CL1}) in Eq. (\ref{system_psi1})
yields a single fractional equation with the cubic nonlinearity:%
\begin{equation}
-D_{1}\left( -\frac{\partial ^{2}}{\partial x^{2}}\right) ^{\alpha /2}\psi
_{1}-\beta _{1}\psi _{1}-\frac{1}{2Q}\left\vert \psi _{1}\right\vert
^{2}\psi _{1}=0.  \label{CL2}
\end{equation}%
Obviously, Eq. (\ref{CL2}) with $\alpha \geq 1$ and $Q<0$ has solitons
solutions, without any threshold (critical) value $N_{\mathrm{cr}}$ of the
norm (\ref{N}) necessary for the existence of the solitons [i.e., $N_{%
\mathrm{cr}}(Q<0,\alpha \geq 1)=0$]. On the contrary, Eq. (\ref{CL2}) with $%
Q>0$ has no bright-soliton solutions for any value of LI $\alpha $. At $Q<0$
and $\alpha <1$, the situation is more complex, because the $z$-dependent
version of Eq. (\ref{CL2}) gives rise to the supercritical collapse in that
case.

It is relevant to note that, for $\beta _{1}\rightarrow +0$, it follows from
Eqs. (\ref{CL2}) and (\ref{CL1}) that the width $L$ and amplitudes $%
A_{1,2}\equiv \left( \left\vert \psi _{1,2}(x)\right\vert \right) _{\max }$
of the corresponding broad solitons asymptotically scale as
\begin{equation}
L\sim \beta _{1}^{-1/\alpha },A_{1}\sim \sqrt{\beta _{1}},A_{2}\sim \beta
_{1},  \label{scaling}
\end{equation}%
hence the power (\ref{N}) scales as%
\begin{equation}
N\sim \beta _{1}^{1-1/\alpha }.  \label{N-scaling}
\end{equation}%
Numerical results for the system with $Q=-1$, presented below in Figs. \ref{figure1}(a) and
\ref{figure2}(a), corroborate the asymptotic relation (\ref{N-scaling}) for $\alpha >1$,
while for $\alpha <1$ the numerical solution of
the full system of equations (\ref{system_psi1}) and (\ref{system_psi2})
demonstrates that the solitons exist only above a finite threshold
(critical) value of the norm, $N>N_{\mathrm{cr}}$, hence CL does not apply
to the case of $\alpha <1$.

In the case of $Q=0$, when the CL is not relevant, Eqs. (\ref{system_Psi1})
and (\ref{system_Psi2}) admit \emph{exact} scaling invariance (not
restricted to $\beta _{1}\rightarrow 0$). The exact relation between $L$ and
$\beta _{1}\rightarrow 0$ is the same as its asymptotic counterpart in Eq. (%
\ref{scaling}), while the exact scalings of the amplitude and power are
different:%
\begin{equation}
\left( A_{1,2}(Q=0)\right) _{\mathrm{exact}}\sim \beta _{1},\left(
N(Q=0)\right) _{\mathrm{exact}}\sim \beta _{1}^{2-1/\alpha }.
\label{scaling Q=0}
\end{equation}%
Further, it is worthy to note that, in interval (\ref{no-coll}), relation $%
N\left( \beta _{1}\right) $ in Eq. (\ref{scaling Q=0}) satisfies the
well-known necessary stability condition given by the Vakhitov-Kolokolov
(VK) criterion, $dN/d\beta _{1}>0$ \cite{VK,Berge}. On the other hand,
precisely at $\alpha =1/2$, Eq. (\ref{scaling Q=0}) yields $dN/d\beta _{1}=0$%
, which is a signature of the above-mentioned critical collapse \cite{Berge}.

Below, scaling is used to set $Q=\pm 1$, unless $Q=0$ (the scaling does not
imply that CL is not relevant). Taking this definition into regard, we
conclude that the predictions of CL agree, in particular, with numerical
results displayed in Fig. \ref{figure3}, which shows $N_{\mathrm{cr}}=0$ for
$Q=-1$ and $\alpha \geq 1$, and $N_{\mathrm{cr}}>0$ in the case of $Q=+1$,
in accordance with the fact that CL predicts no solitons with $N\rightarrow
0 $ in the latter case.

\section{Stationary modes}

\label{Sec III}

\subsection{The existence range of solitons}

\begin{figure}[tbp]
\centering\vspace{0cm} \includegraphics[width=1\linewidth]{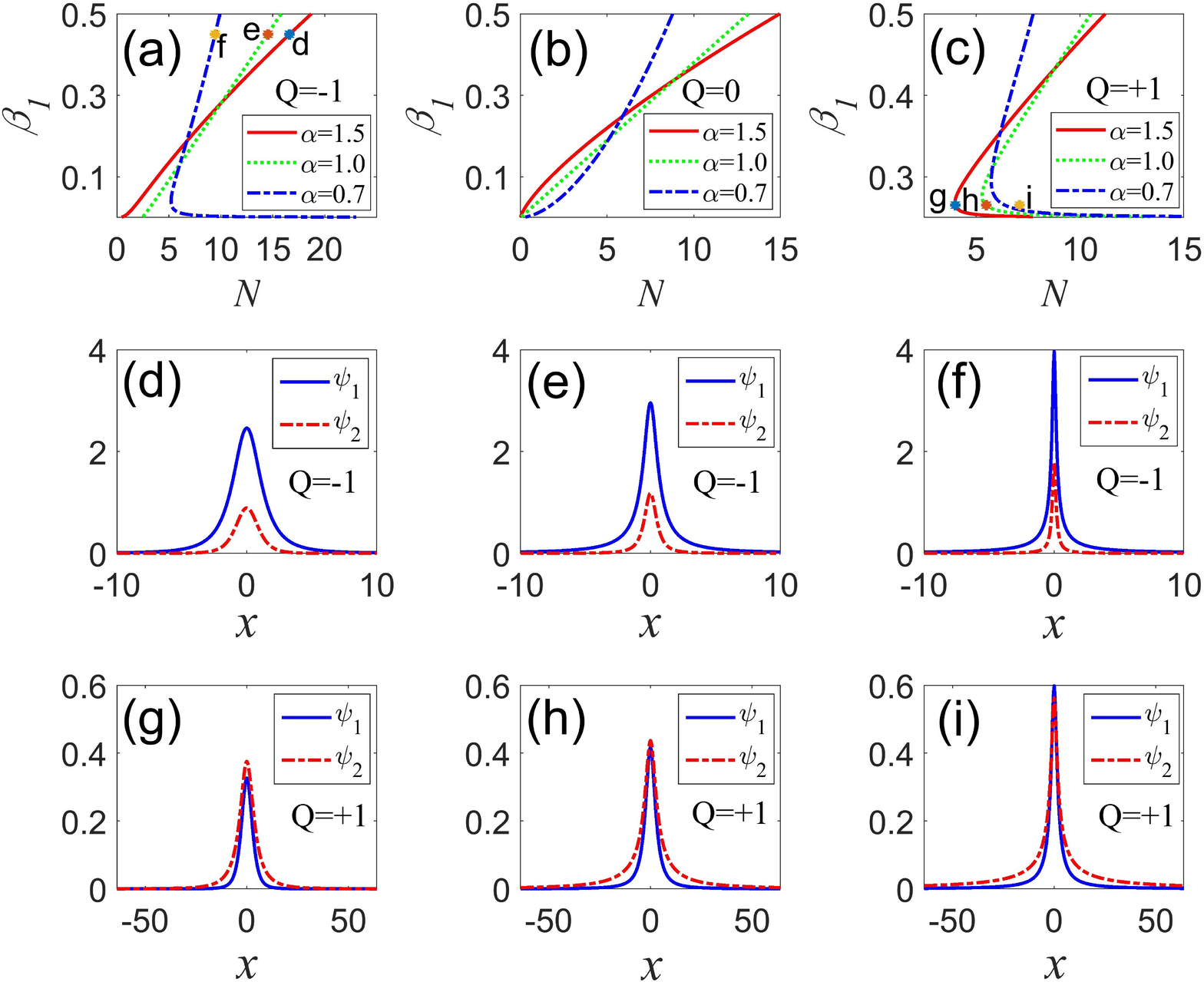}
\vspace{0.0cm}
\caption{The FF propagation constant of the bright solitons, $\protect\beta %
_{1}$, vs. the total power, $N$, for LI values $\protect\alpha =1.5$, $1.0$
and $0.7$, as produced by the numerical solutions of Eqs. (\protect\ref%
{system_psi1}) and (\protect\ref{system_psi2}). Panels (a), (b), and (c)
display the $\protect\beta _{1}(N)$ curves for values of the mismatch
parameter $Q=-1$, $0$, and $+1$, respectively, and for the different values
of LI, as indicated in the figures. The respective soliton families are
stable when they satisfy the the VK criterion, $d\protect\beta _{1}/dN>0$.
Panels (d), (e), and (f) display examples of stable bright solitons with the
FF and SH components $\protect\psi _{1}\left( x\right) $ and $\protect\psi %
_{2}\left( x\right) $ (blue solid and red dashed-dotted lines,
respectively), for a fixed value of $\protect\beta _{1}=0.45$ and mismatch $%
Q=-1$. They correspond to points marked by stars and the same labels, d, e,
and f, in panel\ (a), which identify the respective values of LI. Panels
(g), (h), and (i) display examples of bright solitons for a fixed value of $%
\protect\beta _{1}=0.265$ and mismatch $Q=+1$. They correspond to points
marked by stars and the same labels, g, h, and i, in panel\ (c), which
identify the respective values of LI.}
\label{figure1}
\end{figure}

Numerical solutions of Eqs. (\ref{system_psi1}) and (\ref{system_psi2}) for
stationary two-component solitons, with the FF and SH propagation constants $%
\beta _{1}$ and $\beta _{2}\equiv 2\beta _{1}$, were produced by means of
the Newton-conjugate-gradient method \cite{YangJK-Book}. The results are
presented separately for different values of the mismatch parameter
normalized as mentioned above, i.e., $Q=0$ and $\pm 1$. By means of
rescaling of coordinate $x$, we set $D_{1}=1/2$. Further, because the
diffraction has a universal form for all wave components, we also set $%
D_{2}=1/2$.

The solutions were constructed in interval (\ref{no-coll}) in which, as
mentioned above, the system is free of collapsing. Soliton families are
characterized by dependences $N(\beta _{1})$ for power (\ref{N}), which are
displayed for three characteristic LI values, $\alpha =1.5,$ $1.0,$ and $0.7$%
, and for the negative, zero, and positive mismatch, $Q=-1$, $0$, and $1$,
in Figs. \ref{figure1}(a,b,c).

First, we note that, as predicted above by CL in the case of $Q=-1$ and $%
\alpha >1$, the solitons exist in Fig. \ref{figure1}(a) (actually, for $%
\alpha =1.5$) for all values of $N>0$, i.e., there is no threshold value of
the norm necessary for their existence. Also in agreement with CL, this
soliton family satisfies the VK criterion, $d\beta _{1}/dN>0$ (and it is
indeed a completely stable family). Furthermore, the segment of the $\beta
_{1}(N;\alpha =1.5)$ curve at $\beta _{1}\rightarrow 0$ is consistent with
the asymptotic relation (\ref{scaling}), which predicts, in this case, $%
\beta _{1}\sim N^{3}$ (precise comparison with the latter prediction is
difficult, as one should use an extremely broad integration domain to
produce accurate results for very broad solitons in the limit of $\beta
_{1}\rightarrow 0$).

For $\alpha =1.5$ and $1.0$, the propagation constant $\beta _{1}$ is a
monotonously growing function of the power in Fig. \ref{figure1}(a) for $%
Q=-1 $, while the curve for $\alpha =0.7$ is divided in two branches, with
positive and negative slopes, $d\beta _{1}/dN>0$ and $d\beta _{1}/dN<0$,
indicating that the bright solitons exist at $N>N_{\mathrm{cr}}$. Only the
branch with $d\beta _{1}/dN>0$ may be stable, according to the VK criterion.

Further, the $\beta _{1}(N)$ dependences shown for $Q=0$ in Fig. \ref%
{figure1}(b) fully corroborate the exact relation (\ref{scaling Q=0}). For
example, for $\alpha =1$ the dependence is precisely linear, as predicted by
Eq. (\ref{scaling Q=0}), \textit{viz}., $\beta _{1}=0.038N$, where the
proportionality coefficient is extracted from the numerical data.

Next, Fig. \ref{figure1}(c) shows that all $\beta _{1}(N)$ curves for $Q=+1$
are divided into branches with positive and negative slopes, demonstrating
the existence of a finite threshold (critical) value, $N_{\mathrm{cr}}>0$.
All the turning points in Fig. \ref{figure1}(c), with $N=N_{\mathrm{cr}}$,
correspond to very close values of the propagation constant, $\beta
_{1}\simeq 0.27$. This value can be readily explained by noting that the
full coefficient $(Q-4\beta _{1})$ in front of the linear term in Eq. (\ref{system_psi2}) becomes
negative at $\beta _{1}>Q/4\equiv 0.25$ (recall $Q=1$
is fixed in this case) and, accordingly, Eq. (\ref{CL2}) with $Q$ replaced
by the full coefficient, $(Q-4\beta _{1})$, suggests that solitons may exist
at $\beta _{1}>\left( \beta _{1}\right) _{\mathrm{cr}}\approx 0.25$, even if
CL does not exactly apply for relatively small values of $\left\vert
Q-4\beta _{1}\right\vert $.

Figures \ref{figure1}(d,e,f) show soliton shapes for components $\psi _{1}$
and $\psi _{2}$ at $Q=-1$ with a fixed propagation constant, $\beta
_{1}=0.45 $, for the same values of LI as in Fig. \ref{figure1}(a). The
amplitudes of $\psi _{1}$ and $\psi _{2}$ increase, while the widths of the
profiles shrink, as the value of LI decreases, as smaller LI\ makes the
diffraction weaker, hence the soliton needs to be narrower, to keep the
balance with the nonlinearity. Note that the amplitude and width of the $%
\psi _{2}$ component are essentially smaller than in the $\psi _{1}$
components, which is a natural \textquotedblleft remnant" of CL, see Eq. (%
\ref{CL1}), even if CL does not directly apply for relatively large values
of $\beta _{1}$. Similarly,{\LARGE \ }Figs. \ref{figure1}(g), \ref{figure1}%
(h) and \ref{figure1}(i) show the solitons at $Q=+1$ with fixed $\beta
_{1}=0.265$ (close to $N=N_{\mathrm{cr}}$) for the same values of LI as in
Fig. \ref{figure1}(c), where the shapes of components $\psi _{1}$ and $\psi
_{2}$ are close, unlike the case of $Q=-1$.

\begin{figure}[tbp]
\centering\vspace{0cm} \includegraphics[width=1\linewidth]{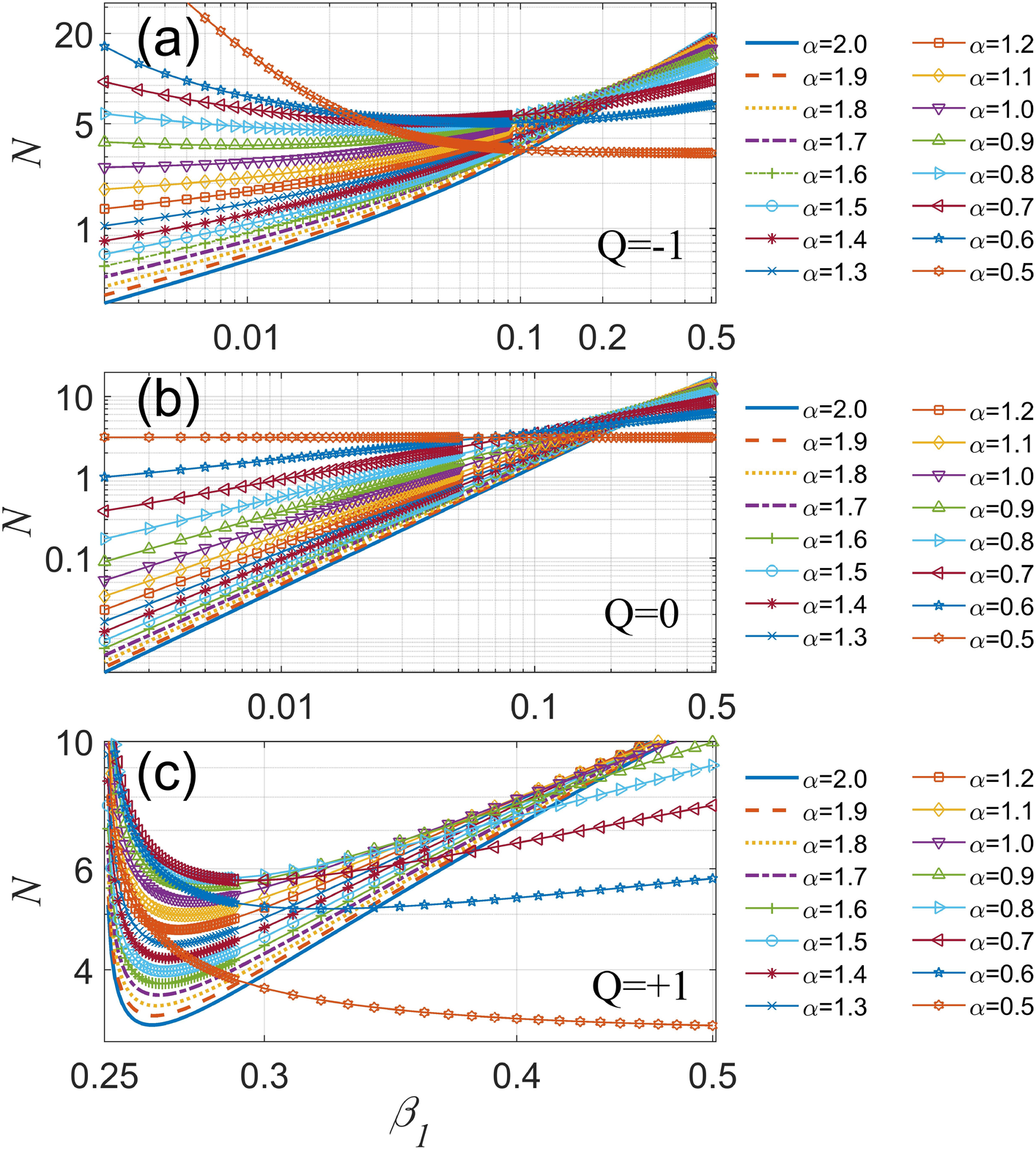}
\vspace{0.0cm}
\caption{Power curves $N\left( \protect\beta _{1}\right) $ for soliton
families with different values of LI $\protect\alpha $ and different
mismatch parameters: (a) $Q=-1$, (b) $Q=0$, and (c) $Q=+1$. Note that the
logarithmic scales are used for $N$ and $\protect\beta _{1}$ in panels (a)
and (b), while (c) is plotted using ordinary scales.}
\label{figure2}
\end{figure}

\subsection{Stability of the solitons}

As mentioned above, the necessary stability condition for the families of
stationary solitons is given by the VK criterion, $d\beta _{1}/dN>0$. To
summarize these results, a large number of $N\left( \beta _{1}\right) $
curves are collected, for $Q=-1$, $0$, and $+1$, in Figs. \ref{figure2}(a),
(b), and (c), respectively. Each panel includes $16$ curves for values of LI
covering the interval (\ref{no-coll}). The range of values of the propagation
constant is $0.002\leq \beta _{1}\leq 0.5$ in Figs. \ref{figure2}(a) and \ref%
{figure2}(b) for $Q=-1$ and $0$, and is $0.0251\leq \beta _{1}\leq 0.5$
for $Q=1$ in Figs. \ref{figure2}(c). The size of the computation domain is $%
X=16384$ for $0.7\leq \alpha \leq 2.0$, $X=131072$ for $\alpha =0.6$, and $%
X=409600$ for $\alpha =0.5$ in Figs. \ref{figure2}(a) and \ref{figure2}(b).
Further, it is $X=16384$ for $0.9\leq \alpha \leq 2.0$ and $X=20480$ for $%
0.5\leq \alpha \leq 0.8$ in Fig. \ref{figure2}(c). For very small values of
the propagation constant $\beta _{1}$, the results may be affected by the
fact that the characteristic width of the solitons, estimated as $\sim \beta
_{1}^{-1/\alpha }$, becomes comparable to the size of the integration domain.

As an extension of the trend observed above in Fig. \ref{figure1}(a), all
the $N\left( \beta _{1}\right) $ curves in Fig. \ref{figure2}(a) for $Q=-1$
show the monotonous growth in the range of $1<\alpha \leq 2$, hence they all
meet the VK criterion. Also in agreement with that trend, at $0.5<\alpha
\leq 1$, there is a finite critical (threshold) value $N_{\mathrm{cr}}$,
which divides the power curves in VK-stable and unstable branches, with $%
dN/d\beta _{1}>0$ and $dN/d\beta _{1}<0$, respectively. In particular, in
the limit of $\alpha =0.5$, the dependence $N\left( \beta _{1}\right) $ is
monotonously decreasing, hence all the solitons are unstable in this limit,
being subject to the onset of the critical collapse, as indicated by the
asymptotically flat dependence for relatively large values of $\beta
_{1}$.

The trend suggested above by Fig. \ref{figure1}(b) for $Q=0$ is confirmed in
detail by Fig. \ref{figure2}(b): all the corresponding $N\left( \beta
_{1}\right) $ curves are monotonously growing at all values of $\beta _{1}$
(clearly suggesting $N_{\mathrm{cr}}=0$) in the entire interval (\ref%
{no-coll}), in accordance with Eq. (\ref{scaling Q=0}), thus providing the
VK stability of all solitons in this interval. In the limit case of $%
\alpha =0.5$, the $N\left( \beta _{1}\right) $ dependence becomes flat, also
in agreement with Eq. (\ref{scaling Q=0}).

The detailed results for $Q=+1$ are collected in Fig. \ref{figure2}(c). In
this case, all curves, except for the one for $\alpha =0.5$, are divided in
VK-stable and unstable branches, separated by $N=N_{\mathrm{cr}}$. In the
limit case of $\alpha =0.5$, the entire $N\left( \beta _{1}\right) $ curve
features a negative slope, i.e., the instability. Note that the unstable
branch, with $dN/d\beta _{1}<0$, persists in the opposite limit of $\alpha
=2 $, which corresponds to the usual $\chi ^{(2)}$ system with
non-fractional diffraction. It is easy to check that, in this limit, the
unstable branch precisely coincides with the previously known one discovered
by the analysis of the usual $\chi ^{(2)}$ system \cite{SH1,SH-review1}.

On the contrary to the cases of $Q=-1$ and $0$, where very broad solitons
exist with arbitrarily small values of the propagation constant $\beta _{1}$
[see Eq. (\ref{scaling})] in the case of $Q=+1$ the values are limited,
according to the form of the linear terms in Eq. (\ref{system_psi2}), to $%
\beta _{1}>1/4$ (as was mentioned above). As well as in the limit of $\beta
_{1}\rightarrow 0$ for $Q=-1$ and $0$, the limit of $\beta
_{1}-1/4\rightarrow 0$ for $Q=+1$ naturally corresponds to solitons with
diverging width $\sim \left( \beta _{1}-1/4\right) ^{-1/\alpha }$.

\begin{figure}[tbp]
\centering\vspace{0cm} \includegraphics[width=1\linewidth]{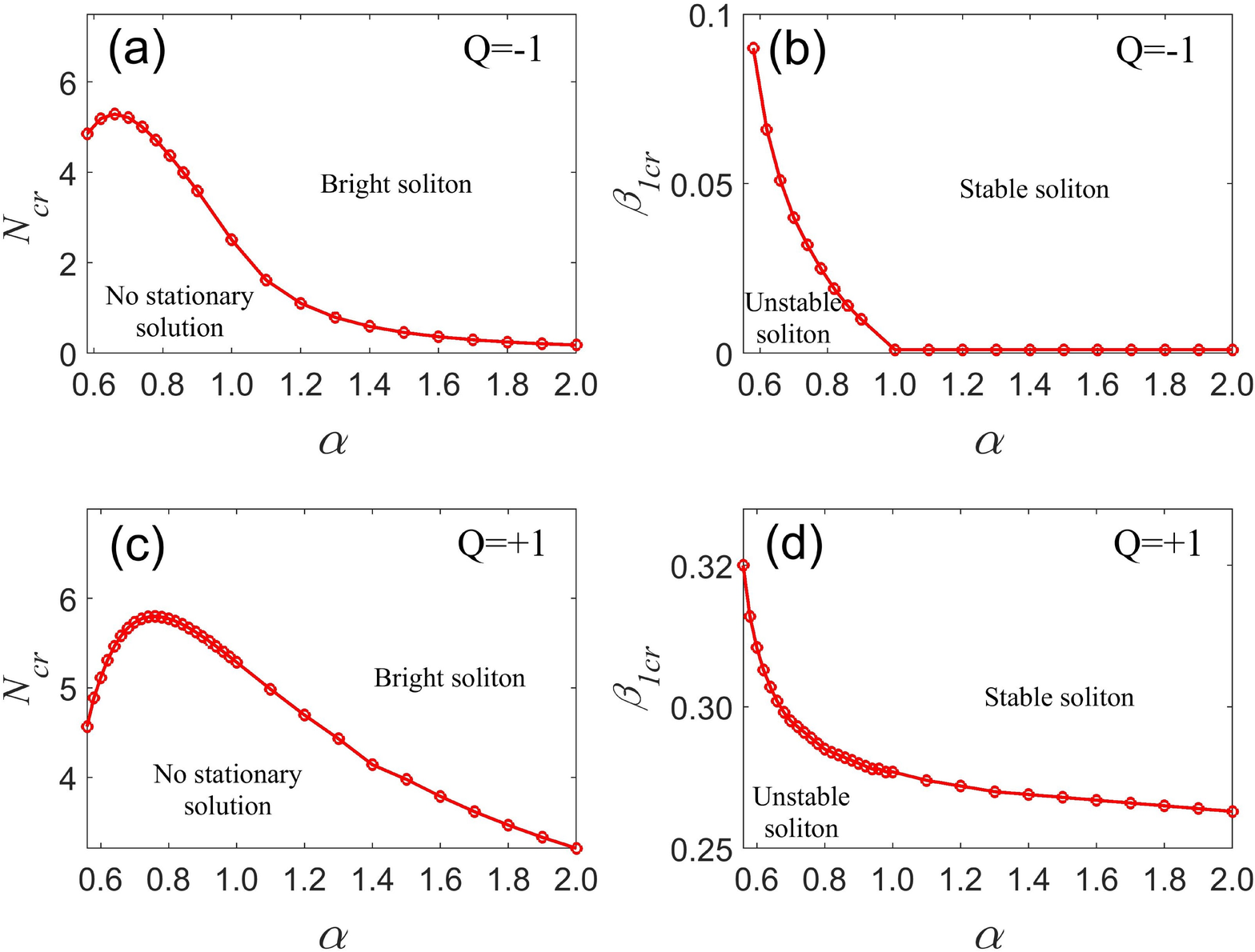}
\vspace{0.0cm}
\caption{Domains of the existence, stability, and instability of the
solitons, divided by the power threshold (critical value) $N_{\mathrm{cr}}$
(a,c) and the respective critical value $\protect\beta _{1\mathrm{cr}}$ of
the propagation constant (b,d) for mismatch parameters $Q=-1$ (a,b) and $%
Q=+1 $ (c,d).}
\label{figure3}
\end{figure}

The numerically found dependence of the critical (threshold) power $N_{%
\mathrm{cr}}$ on LI in the range of $0.5<\alpha \leq 2$ is displayed, for
mismatch $Q=-1$ and $+1$, in Figs. \ref{figure3}(a) and (c). As mentioned
above, the solitons exist at $N>N_{\mathrm{cr}}$, there being no stationary
soliton solutions at $N<N_{\mathrm{cr}}$. The related dependence of the
critical propagation constant $\beta _{1\mathrm{cr}}\equiv \beta _{1}(N_{%
\mathrm{cr}})$ on LI, are boundaries between stable solitons at $\beta
_{1}>\beta _{1\mathrm{cr}}$ and unstable ones at $\beta _{1}<\beta _{1%
\mathrm{cr}}$ for given $\alpha $, as shown in Figs. \ref{figure3}(b) and
(d). These figures corroborate the above-mentioned conclusion, that, for $%
Q=-1$, stable solitons exist for any $N>0$ in the range of $1<\alpha \leq 2$%
, and for $N>N_{\mathrm{cr}}>0$ in the range of $0.5<\alpha \leq 1$; for $%
Q=+1$, the stable solitons exist only above the threshold value, at $N>N_{%
\mathrm{cr}}$, in the entire interval (\ref{no-coll}). On the other hand,
for $Q=0$ all the solitons are stable, as inferred above, for any $N>0$ and for
all values of LI in interval (\ref{no-coll}). Lastly, all the solitons are
unstable at $\alpha =0.5$, for $Q=\pm 1$ and $0$.

\section{Verification of the soliton's stability and instability}

\label{Sec IV}

\subsection{Perturbed solutions}

Predictions for the (in)stability of the solitons, produced above on the
basis of the VK criterion, which, strictly speaking, is a necessary one but
not sufficient, were corroborated by the analysis of linearized equations
for small perturbations and direct simulations of the perturbed evolution.
First, weakly perturbed solutions were looked as

\begin{eqnarray}
\Psi _{1} &=&e^{i\beta _{1}z}\left[ \psi _{1}\left( x\right) +u_{1}\left(
x\right) e^{\delta z}+u_{2}^{\ast }\left( x\right) e^{\delta ^{\ast }z}%
\right] ,  \notag \\
\Psi _{2} &=&e^{i\beta _{2}z}\left[ \psi _{2}\left( x\right) +v_{1}\left(
x\right) e^{\delta z}+v_{2}^{\ast }\left( x\right) e^{\delta ^{\ast }z}%
\right] ,  \label{Perturbation}
\end{eqnarray}%
where $\psi _{1,2}$ represents the unperturbed soliton, defined as per Eq. (%
\ref{Psi1Psi2}), while $u_{1,2}$ and $v_{1,2}$ are the small perturbations
with the respective complex eigenvalue $\delta $. The substitution of
expression (\ref{Perturbation}) in Eqs. (\ref{system_Psi1}) and (\ref%
{system_Psi2}) and linearization leads to the system of coupled equations,

\begin{eqnarray}
\delta u_{1} &=&i\left\{ \left[ -\beta _{1}-D_{1}\left( -\frac{\partial ^{2}%
}{\partial x^{2}}\right) ^{\alpha /2}\right] u_{1}+vu_{2}+u^{\ast
}v_{1}\right\} ,  \notag \\
\delta u_{2} &=&i\left\{ -vu_{1}+\left[ \beta _{1}+D_{1}\left( -\frac{%
\partial ^{2}}{\partial x^{2}}\right) ^{\alpha /2}\right] u_{2}-uv_{2}\right%
\} ,  \notag \\
\delta v_{1} &=&i\left\{ \frac{1}{2}uu_{1}-\left[ \beta _{2}+\frac{D_{2}}{2}%
\left( -\frac{\partial ^{2}}{\partial x^{2}}\right) ^{\alpha /2}-\frac{Q}{2}%
\right] v_{1}\right\} ,  \notag \\
\delta v_{2} &=&i\left\{ -\frac{1}{2}u^{\ast }u_{2}+\left[ \beta _{2}+\frac{%
D_{2}}{2}\left( -\frac{\partial ^{2}}{\partial x^{2}}\right) ^{\alpha /2}-%
\frac{Q}{2}\right] v_{2}\right\} .  \label{Linearization}
\end{eqnarray}%
Equations (\ref{Linearization}) were solved by means of the Fourier
collocation method \cite{YangJK-Book}. The solitons are linearly stable if
all eigenvalues $\delta $ are imaginary, whereas they are unstable in the
case if $\mathrm{Re}(\delta )>0$ exists.

Direct tests of the perturbed stability of the solitons were performed by
simulations of Eqs. (\ref{system_Psi1}) and (\ref{system_Psi2}) with input
taken as per Eq. (\ref{Perturbation}) at $z=0$, using the spectral method
combined with the Runge-Kutta one.

\subsection{Stable and unstable perturbation eigenvalues and direct
propagation}

\begin{figure}[tbp]
\centering\vspace{0cm} \includegraphics[width=1\linewidth]{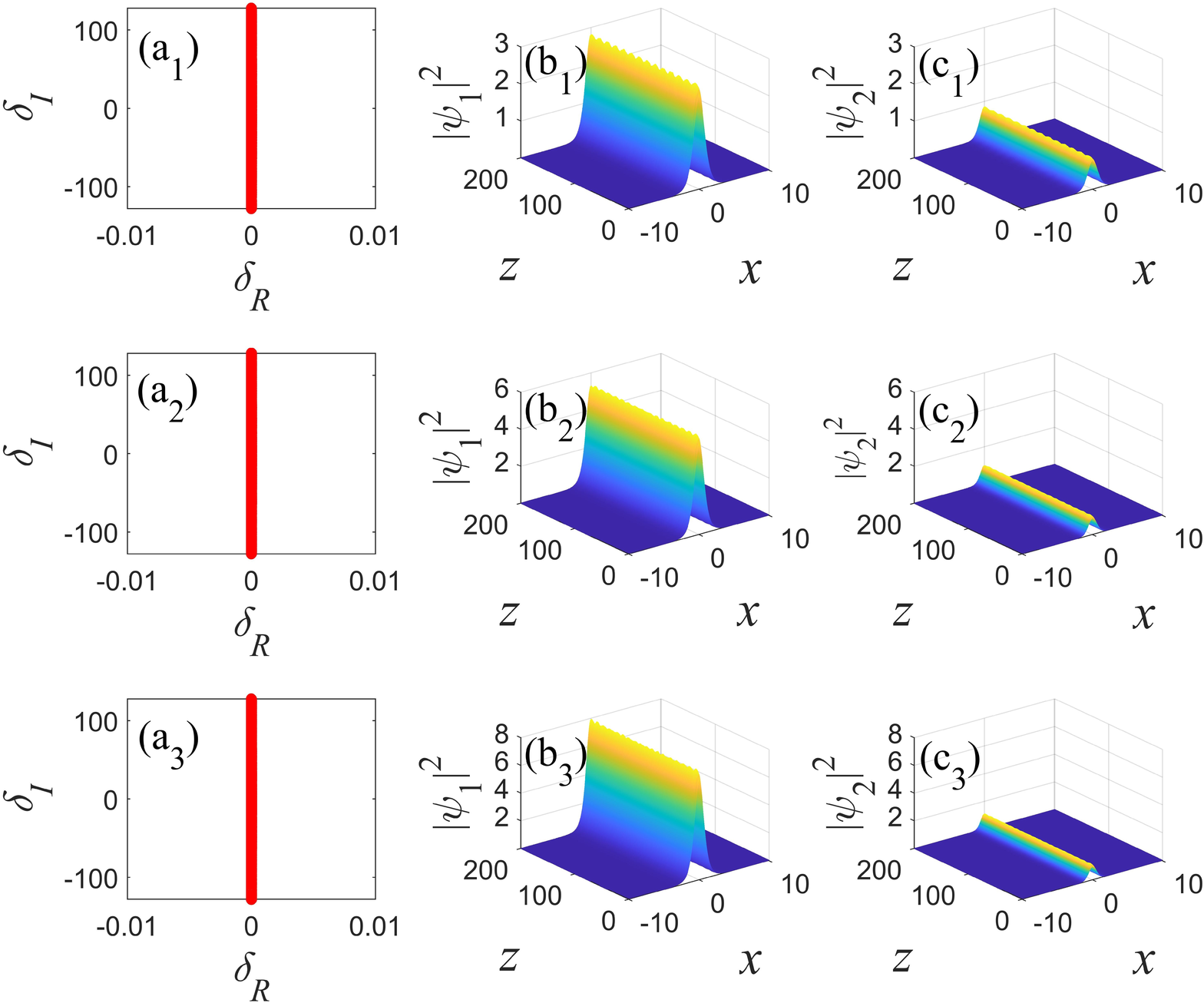}
\vspace{0.0cm}
\caption{Linear-stability spectra for stable solitons (\textrm{a}$_{1}$,
\textrm{a}$_{2}$,\textrm{a}$_{3}$) and the simulated evolution of the FF and
SH components initiated with $5\%$ random perturbations (\textrm{b}$_{1}$,
\textrm{b}$_{2}$,\textrm{b}$_{3}$) and (\textrm{c}$_{1}$,\textrm{c}$_{2}$,
\textrm{c}$_{3}$). Values of the mismatch parameter corresponding to the
top, middle, and bottom rows are $Q=+1$, $0$, and $-1$, respectively. Other
parameters are fixed as $\protect\beta _{1}=0.5$ and $\protect\alpha =1.5$.}
\label{figure4}
\end{figure}

We have corroborated the stability of the solitons, as predicted above by
the VK criterion, by results of the linear-stability analysis and direct
simulations. Examples are demonstrated in Fig. \ref{figure4} for $\alpha
=1.5 $, $\beta _{1}=0.5$, and $Q=\pm 1$ and $0$. It is seen in panels (a$_{1}
$)--(a$_{3}$) that VK-stable solitons are indeed linearly stable. To
simulate the full system of Eqs. (\ref{system_Psi1}) and (\ref{system_Psi2}%
), random perturbations at the $5\%$ amplitude level were added at $z=0$ to
the stationary solitons, as per Eq. (\ref{Perturbation}).

\begin{figure}[tbp]
\centering\vspace{0cm} \includegraphics[width=1\linewidth]{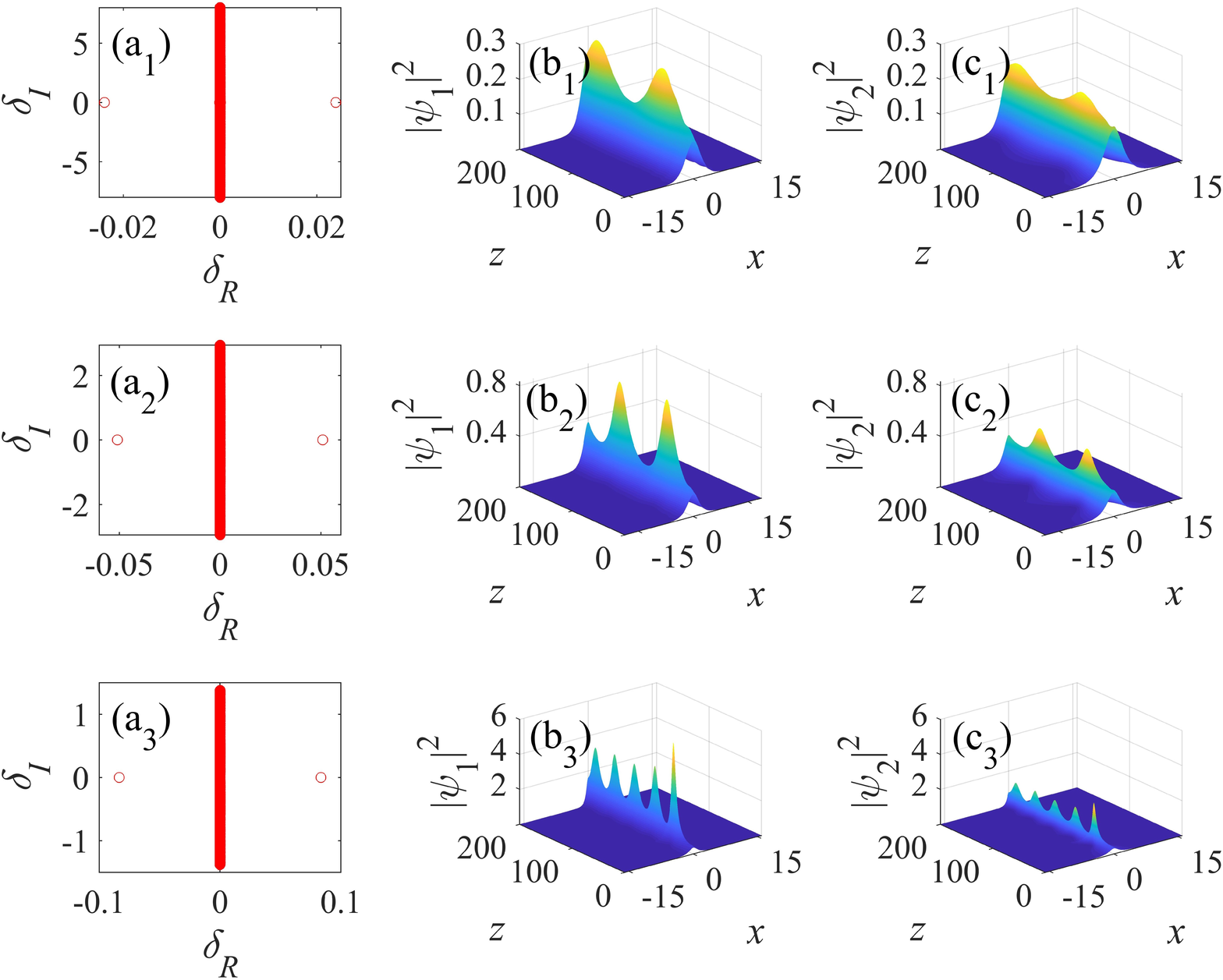}
\vspace{0.0cm}
\caption{The linear-stability spectra ($\mathrm{a}_{1}$,$\mathrm{a}_{2}$,$
\mathrm{a}_{3}$) and perturbed evolution of FF and SH components ($\mathrm{b}
_{1}$,$\mathrm{b}_{2}$,$\mathrm{b}_{3}$) and ($\mathrm{c}_{1}$,$\mathrm{c}_{2}$
,$\mathrm{c}_{3}$) for VK-unstable solitons with mismatch parameter $Q=1$
and propagation constant $\protect\beta _{1}=0.265$. Top, middle, and bottom
rows correspond, respectively, to different LI values, \textit{viz}., $%
\protect\alpha =1.5$, $\protect\alpha =1.0$, and $\protect\alpha =0.7$.}
\label{figure5}
\end{figure}

The VK-predicted instability of the solitons has been verified too. Examples
are displayed for $Q=1$, $\beta _{1}=0.265$ and several different values of
LI in Fig. \ref{figure5}. It is worthy to note that unstable solitons are not fully
destroyed by perturbations, but spontaneously transform into robust
oscillatory states.

\begin{figure}[tbp]
\centering\vspace{0cm} \includegraphics[width=1\linewidth]{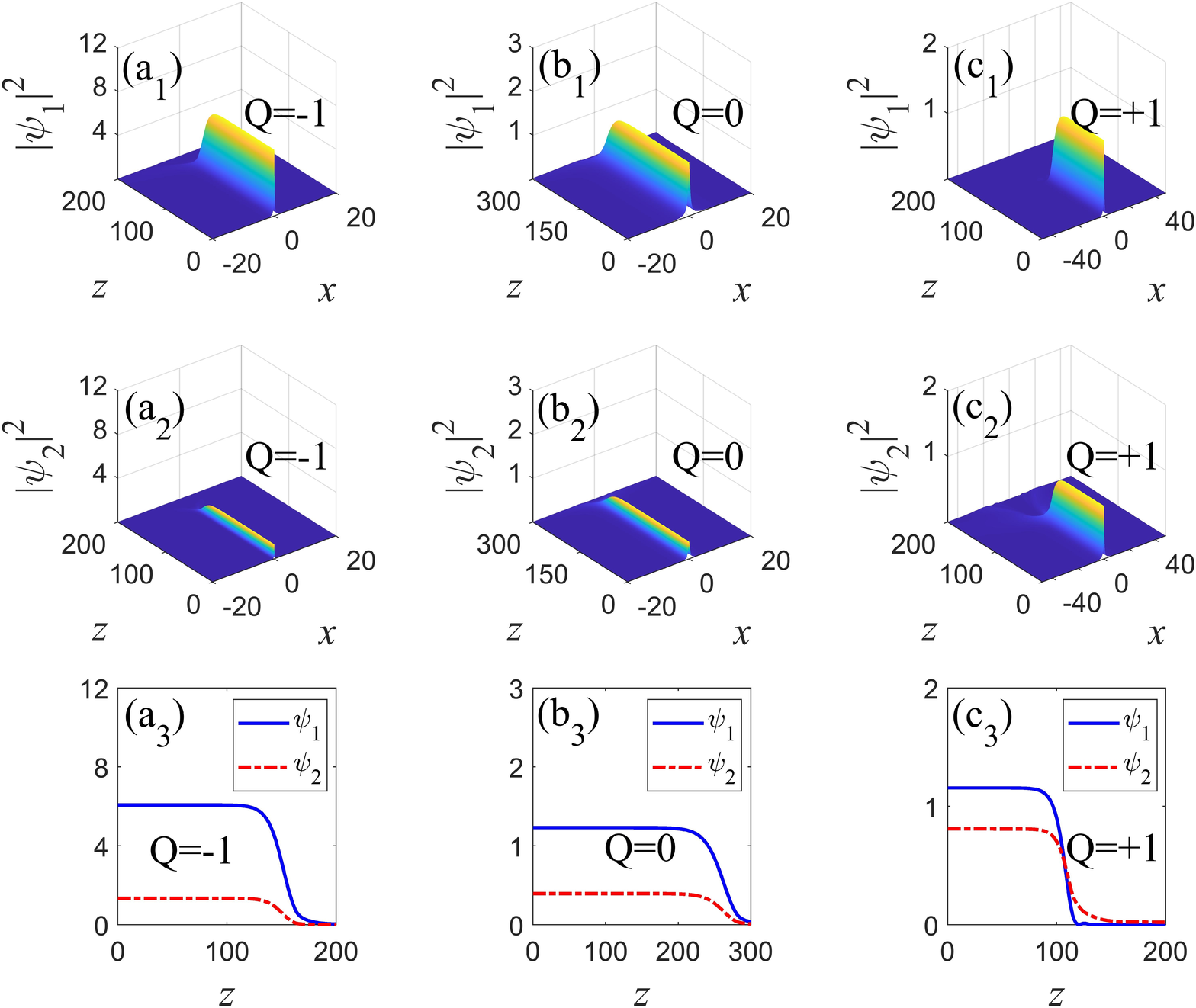}
\vspace{0.0cm}
\caption{The unstable evolution of the FF and SH components of the solitons
at the value of LI $\protect\alpha =0.5$, at which Eqs. (\protect\ref%
{system_Psi1}) and (\protect\ref{system_Psi2}) give rise to the critical
collapse. Panels (a$_{1}$--a$_{3}$), (b$_{1}$--b$_{3}$) display examples of
the spontaneous decay for values of the mismatch $Q=-1$ and $0$,
respectively, with fixed propagation constant $\protect\beta _{1}=0.1$.\
Panels (c$_{1}$--c$_{3}$) display the spontaneous decay for the mismatch $%
Q=+1$ with fixed propagation constant $\protect\beta _{1}=0.265$.}
\label{figure6}
\end{figure}

The instability of the solitons for $\alpha =0.5$ is also corroborated by
direct simulations of Eqs. (\ref{system_Psi1}) and (\ref{system_Psi2}), see
examples in Fig. \ref{figure6} for propagation constant $\beta _{1}=0.1$ and
mismatch parameters $Q=-1$ and $0$ in Figs. \ref{figure6}(a$_{1}$)--(a$_{3}$%
) and \ref{figure6}(b$_{1}$)--(b$_{3}$), respectively, and for propagation
constant $\beta _{1}=0.265$ with mismatch parameter $Q=1$ in Figs. \ref{figure6}(c$_{1}$)--(c$_{3}$).
In these examples, the critical collapse manifests itself
by sudden decay of both the FF and SH components.

\begin{figure}[tbp]
\centering\vspace{0cm} \includegraphics[width=1\linewidth]{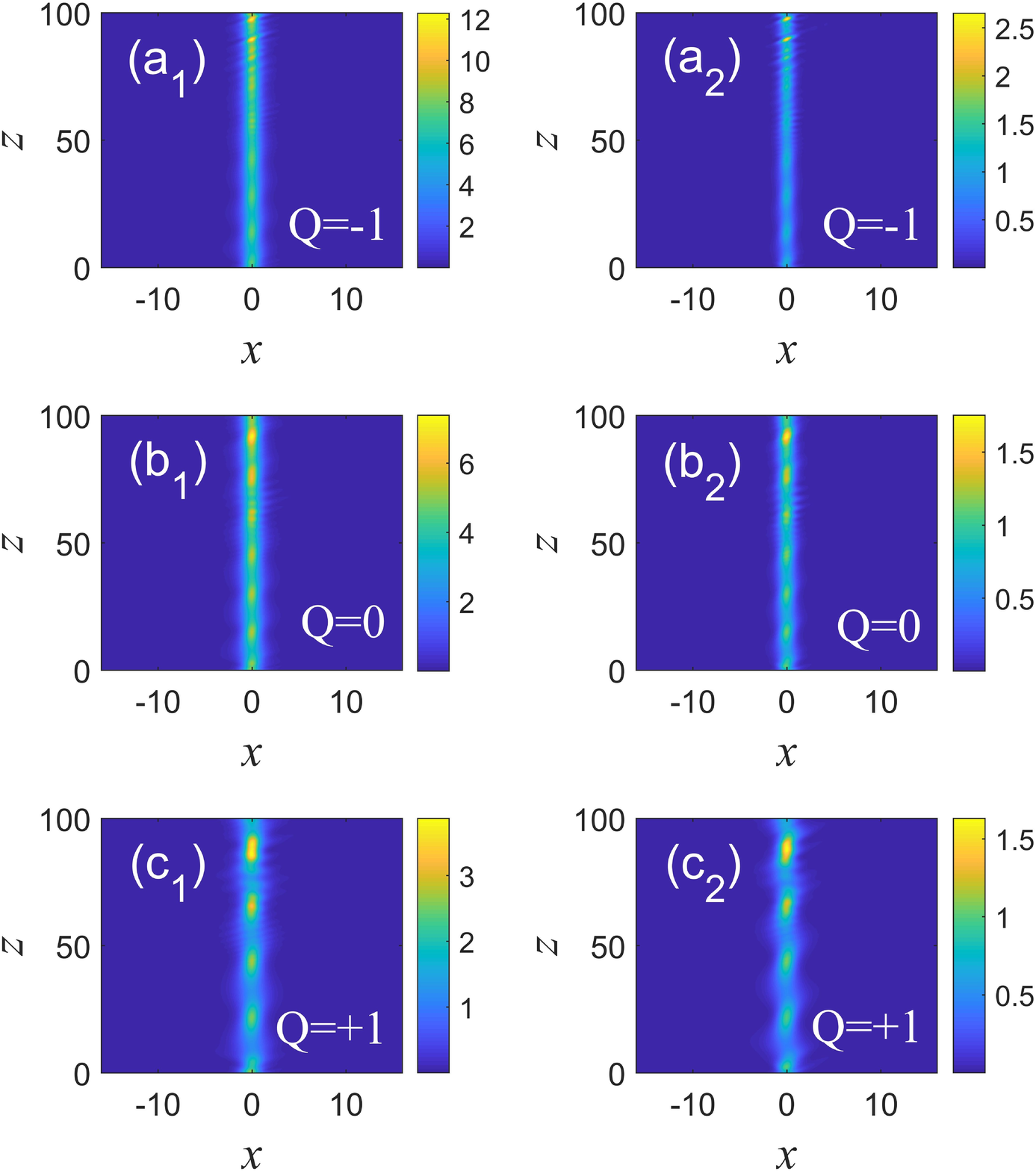}
\vspace{0.0cm}
\caption{The evolution of a soliton initiated by kick $k=0.3$ in Eq. (%
\protect\ref{Psi1_Psi2_k1k2}), displayed by means of local-density maps of
the FF and SH local powers, $\left\vert \protect\psi _{1}(x.z)\right\vert
^{2}$ and $\left\vert \protect\psi _{2}(x.z)\right\vert ^{2}$, in the left
and right panels, respectively. The top, middle, and bottom rows of panels
correspond to the mismatch parameters $Q=-1$, $0$, and $+1$, respectively.
Other parameters are $\protect\beta _{1}=0.5$ and $\protect\alpha =1.5$. In
this case, the kick excites an intrinsic mode in the soliton, but does not
set it in motion.}
\label{figure7}
\end{figure}

\begin{figure}[tbp]
\centering\vspace{0cm} \includegraphics[width=1\linewidth]{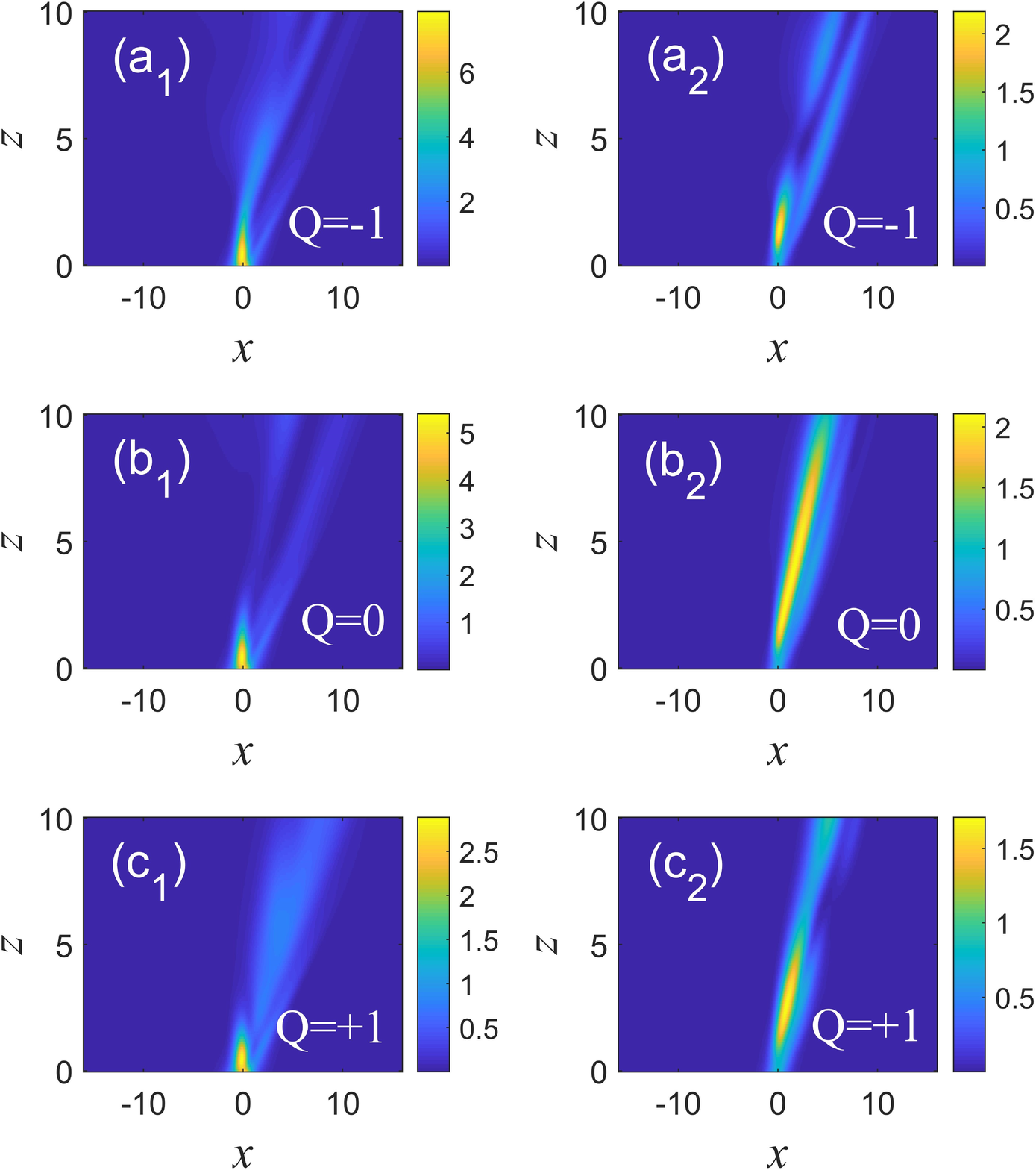}
\vspace{0.0cm}
\caption{The same as in Fig. \protect\ref{figure7}, but for $k=1$. In this
case, the sufficiently strong kick initiates the motion of the soliton,
simultaneously destroying it.}
\label{figure8}
\end{figure}

\subsection{Kicked solitons}

The fractional diffraction breaks the Galilean invariance of underlying
equations (\ref{system_Psi1}) and (\ref{system_Psi2}) \cite{Malomed-Rev},
therefore attempts to construct solutions for \textquotedblleft moving"
solitons (actually, for ones tilted in the spatial domain) is a nontrivial
objective. It may be addressed by taking a stable stationary soliton and
applying transverse kicks $k$ and $2k$ to its FF and SH components, i.e.,
simulating (\ref{system_Psi1}) and (\ref{system_Psi2}) with input
\begin{equation}
\Psi _{1}\left( x\right) =\psi _{1}(x)\exp \left( ikx\right) ,\Psi
_{2}\left( x\right) =\psi _{2}(x)\exp \left( 2ikx\right) .
\label{Psi1_Psi2_k1k2}
\end{equation}%
The simulations demonstrate that, for $k$ small enough, the kick \emph{does
not} initiate motion, in contrast with the commonly known result for the
Galilean-invariant systems, such as Eqs. (\ref{system_Psi1}) and (\ref%
{system_Psi2}) with the non-fractional diffraction, $\alpha =2$. Instead, a
relatively weak kick excites robust intrinsic vibrations in the quiescent
soliton, as shown in Fig. \ref{figure7} for $\alpha =1.5$, $\beta _{1}=0.5$,
and $k=0.3$. It is worthy to note that the internal vibrations are very weak
in the case of $Q=-1$, being much more conspicuous for $Q=+1$. This
observation may be qualitatively explained by the fact that the single
equation (\ref{CL2}), to which the underlying system of Eqs. (\ref%
{system_psi1}) and (\ref{system_psi2}) is reduced by dint of the CL
approximation in the case of $Q=-1$, is close to the usual NLS equation.
Further, it is commonly known that the usual NLS bright solitons do not
support any intrinsic mode, while the usual full $\chi ^{(2)}$ system,
corresponding to $Q=+1$, gives rise to such a mode \cite{SH1,SH-review1}.
While this argument helps to understand the difference between the cases of $%
Q=-1$ and $+1$ in Fig. \ref{figure7}, detailed studies of the intrinsic
modes in solitons generated by the fractional system is a subject for a
separate work, which will be reported elsewhere.

An essentially larger kick is able to set the soliton in motion, but
simultaneously destroying it, as shown in Fig. \ref{figure8}. The
possibility to create stably moving (tilted) solitons remains a subject for
systematic studies. It will also be relevant to accurately identify critical
values of the kick leading to the destruction of the originally quiescent
solitons, and a possible link between the excitation of the above-mentioned
intrinsic mode and eventual destruction.

\section{Conclusion}

\label{Sec V} We have extended the recently introduced concept of solitons
in nonlinear optical waveguides with fractional diffraction for the
second-harmonic-generation system with the $\chi ^{(2)}$ nonlinearity.
Families of fundamental solitons have been constructed for values of L\'{e}vy index between
$\alpha =2$, which corresponds to the normal
non-fractional diffraction, and $\alpha =0.5$, which gives rise to the
critical collapse, in the case of the quadratic nonlinearity, and for
positive, negative, and zero values of the mismatch between the
fundamental-frequency and second-harmonic components of the $\chi
^{(2)}$ system. Some characteristics of the soliton families are found in
the analytical form, using the cascading limit and scaling relations. Through
the computation of stability eigenvalues for small perturbations and direct
simulations of the perturbed evolution of solitons, it has been found that the
Vakhitov-Kolokolov criterion for the soliton families exactly predicts
their stability. In the case of $\alpha >0.5$, unstable solitons
spontaneously transform into robust breathers, while in the case of $\alpha
=0.5$ the critical collapse leads to spontaneous decay of unstable solitons.
The application of a relatively weak transverse kick excites small internal
vibrations in the stable solitons, failing to set them in motion. A stronger
kick initiates the motion (transverse tilt) of the solitons, simultaneously
destroying them.

The analysis reported in this work may be developed in other directions. In
particular, it may be interesting to produce fractional solitons in the
two-dimensional spatial $\chi ^{(2)}$ system, as well as in the
spatiotemporal one.

\section{Acknowledgments}

This work was supported by National Natural Science Foundation of China
(11805141, 62205224), Guangdong Basic and Applied Basic Research Foundation
(2023A1515010865), and the Scientific and Technological Innovation Programs
of Higher Education Institutions in Shanxi (STIP) (2021L401). The work of D.M.
was supported by the Romanian Ministry of Research, Innovation, and
Digitization (PN 23210101/2023). The work of B.A.M. was supported, in part, by the
Israel Science Foundation through the Grant No. 1695/22.

\section*{Declaration of Competing Interest}

The authors declare that they have no competing financial interests or
personal relationships that might influence the work reported in this paper.


\begin{thebibliography}{99}
\bibitem{Laskin1} Laskin N. Fractional quantum mechanics.\ Phys Rev E
2000;62(3):3135--45.

\bibitem{Laskin2} Laskin N. Fractional quantum mechanics and L\'{e}vy path
integrals.\ Phys Lett A 2000;268(4--6):298--305.

\bibitem{Laskin3} Laskin N. Fractional Schr\"{o}dinger equation.\ Phys Rev E
2002;66(5):056108.

\bibitem{Book-Laskin} Laskin N. Fractional quantum mechanics. World
Scientific Publishing Co. Pte. Ltd.; 2018.

\bibitem{Levy-Crystal} Stickler BA. Potential condensed-matter realization
of space fractional quantum mechanics: The one-dimensional L\'{e}vy crystal.
Phys Rev E 2013;88(1):012120.

\bibitem{Polariton-conden} Pinsker F, Bao W, Zhang Y, Ohadi H, Dreismann A,
Baumberg JJ. Fractional quantum mechanics in polariton condensates with
velocity-dependent mass. Phys Rev B 2015;92(19):195310.

\bibitem{FD-Optics1} Longhi S. Fractional Schr\"{o}dinger equation in
optics.\ Opt Lett 2015;40(6):1117--20.

\bibitem{fNLSE1} Zhong W, Beli\'{c} MR, Zhang Y. Accessible solitons of
fractional dimension. Ann Phys 2016;368:110--6.

\bibitem{fNLSE4} Zhong W, Beli\'{c} MR, Malomed BA, Zhang Y, Huang T.
Spatiotemporal accessible solitons in fractional dimensions. Phys Rev E
2016;94(1):012216.

\bibitem{fNLSE2} Huang C, Dong L. Gap solitons in the nonlinear fractional
Schr\"{o}dinger equation with an optical lattice. Opt Lett
2016;41(24):5636--9.

\bibitem{fNLSE7} Huang C, Dong L. Composition relation between nonlinear
Bloch waves and gap solitons in periodic fractional systems. Materials
2018;11(7):1134.

\bibitem{fNLSE8} Zeng LW, Beli\'{c} MR, Mihalache D, Shi J, Li J, Li S, Lu
X, Cai Y, Li J. Families of gap solitons and their complexes in media with
saturable nonlinearity and fractional diffraction. Nonlinear Dyn
2022;108(2):1671--80.

\bibitem{fNLSE5} Xiao J, Tian Z, Huang C, Dong L. Surface gap solitons in a
nonlinear fractional Schr\"{o}dinger equation Opt Express
2018;26(3):2650--8.

\bibitem{fNLSE11} Huang C, Dong L. Dissipative surface solitons in a
nonlinear fractional Schr\"{o}dinger equation. Opt Lett 2019;44(22):5438--41.

\bibitem{fNLSE9} Yao X, Liu X. Off-site and on-site vortex solitons in
space-fractional photonic lattices. Opt Lett 2018;43(23):5749--52.

\bibitem{fNLSE10} Zeng LW, Zeng JH. One-dimensional solitons in fractional
Schr\"{o}dinger equation with a spatially periodical modulated nonlinearity:
nonlinear lattice. Opt Lett 2019;44(11):2661--4.

\bibitem{fNLSE3} Chen MN, Zeng SH, Lu DQ, Hu W, Guo Q. Optical solitons,
self-focusing, and wave collapse in a space-fractional Schr\"{o}dinger
equation with a Kerr-type nonlinearity. Phys Rev E 2018;98(2):022211.

\bibitem{fNLSE6} Chen MN, Guo Q, Lu DQ, Hu W. Variational approach for
breathers in a nonlinear fractional Schr\"{o}dinger equation. Commun
Nonlinear Sci Numer Simulat 2019;71:73--81.

\bibitem{fNLSE12} Molina MI. The fractional discrete nonlinear Schr\"{o}%
dinger equation. Phys Lett A 2020;384(8):126180.

\bibitem{fNLSE13} Qiu YL, Malomed BA, Mihalache D, Zhu X, Zhang L, He YJ.
Soliton dynamics in a fractional complex Ginzburg-Landau model. Chaos
Solitons Fractals 2020;131:109471.

\bibitem{fNLSE14} Li PF, Malomed BA, Mihalache D. Symmetry breaking of
spatial Kerr solitons in fractional dimension. Chaos Solitons Fractals
2020;132:109602.

\bibitem{fNLSE15} Wang BH, Lu PH, Dai CQ, Chen YX. Vector optical soliton
and periodic solutions of a coupled fractional nonlinear Schr\"{o}dinger
equation. Results Phys 2020;17:103036.

\bibitem{fNLSE16} Chen JB, Zeng JH. Spontaneous symmetry breaking in purely
nonlinear fractional systems. Chaos 2020;30(6):063131.

\bibitem{fNLSE17} Qiu YL, Malomed BA, Mihalache D, Zhu X, Peng X, He Y.
Stabilization of single- and multi-peak solitons in the fractional nonlinear
Schr\"{o}dinger equation with a trapping potential. Chaos Solitons Fractals
2020;140:110222.

\bibitem{fNLSE18} Li PF, Malomed BA, Mihalache D. Vortex solitons in
fractional nonlinear Schr\"{o}dinger equation with the cubic-quintic
nonlinearity. Chaos Solitons Fractals 2020;137:109783.

\bibitem{fNLSE19} Wang Q, Zhang LL, Malomed BA, Mihalache D, Zeng LW.
Transformation of multipole and vortex solitons in the nonlocal nonlinear
fractional Schr\"{o}dinger equation by means of L\'{e}vy-index management.
Chaos Solitons Fractals 2022;157:111995.

\bibitem{fNLSE20} Zeng LW, Zhu YL, Malomed BA, Mihalache D, Wang Q, Long H,
Cai Y, Lu XW, Li JZ. Quadratic fractional solitons. Chaos Solitons Fractals
2022;154:111586.

\bibitem{fNLSE-PT1} Zhong WP, Beli\'{c} MR, Zhang YQ. Fractional dimensional
accessible solitons in a parity-time symmetric potential. Ann Phys
2018;530(2):1700311.

\bibitem{fNLSE-PT2} Dong LW, Huang CM. Double-hump solitons in fractional
dimensions with a $\mathcal{PT}$-symmetric potential. Opt Express
2018;26(8):10509--18.

\bibitem{fNLSE-PT3} Huang CM, Deng HY, Zhang WF, Ye FW, Dong LW. Fundamental
solitons in the nonlinear fractional Schr\"{o}dinger equation with a
PT-symmetric potential. Europhys Lett 2018;122(2):24002.

\bibitem{fNLSE-PT4} Yao XK, Liu XM. Solitons in the fractional Schr\"{o}%
dinger equation with parity-time-symmetric lattice potential. Photonics Res
2018;6(9):875--9.

\bibitem{fNLSE-PT5} Dong LW, Huang CM. Vortex solitons in fractional systems
with partially parity-time-symmetric azimuthal potentials. Nonlinear Dyn
2019;98(2):1019--28.

\bibitem{fNLSE-PT6} Zhu X, Yang FW, Cao SL, Xie JQ, He YJ. Multipole gap
solitons in fractional Schr\"{o}dinger equation with parity-time-symmetric
optical lattices. Opt Express 2020;28(2):1631--9.

\bibitem{fNLSE-PT7} Li PF, Malomed BA, Mihalache D. Symmetry-breaking
bifurcations and ghost states in the fractional nonlinear Schr\"{o}dinger
equation with a PT-symmetric potential. Opt Lett 2021;46(13):3267--70.

\bibitem{fNLSE-PT8} Wu ZK, Yang KB, Ren XJ, Li P, Wen F, Gu YZ, Guo LJ.
Conical diffraction modulation in fractional dimensions with a PT-symmetric
potential. Chaos Solitons Fractals 2022;164:112631.

\bibitem{CFNLSE1} Zeng LW, Zeng JH. Fractional quantum couplers. Chaos
Solitons Fractals 2020;140:110271.

\bibitem{CFNLSE2} Zeng LW, Shi JC, Lu XW, Cai Y, Zhu QF, Chen HY, Long H, Li
JZ. Stable and oscillating solitons of $\mathcal{PT}$-symmetric couplers
with gain and loss in fractional dimension. Nonlinear Dyn
2021;103(2):1831--40.

\bibitem{CFNLSE3} Zeng LW, Beli\'{c} MR, Mihalache D, Wang Q, Chen JB, Shi
JC, Cai Y, Lu XW, Li JZ. Solitons in spin-orbit-coupled systems with
fractional spatial derivatives. Chaos Solitons Fractals 2021;152:111406.

\bibitem{DW-FNLSE} Kumar S, Li PF, Malomed BA. Domain walls in fractional
media. Phys Rev E 2022;106(5):054207.

\bibitem{Malomed-Rev} Malomed BA. Optical solitons and vortices in
fractional media: A mini-review of recent results. Photonics 2021;8(9):353.

\bibitem{Torruellas} Torruellas WE, Wang Z, Hagan DJ, VanStryland EW,
Stegeman GI, Torner L, Menyuk CR. Observation of two-dimensional spatial
solitary waves in a quadratic medium. Phys Rev Lett 1995;74(25):5036--9.

\bibitem{SH1} Pelinovsky DE, Buryak AV, Kivshar YS. Instability of solitons
governed by quadratic nonlinearities. Phys Rev Lett 1995;75(4):591--5.

\bibitem{SH2} Buryak AV, Kivshar YS. Solitons due to second harmonic
generation. Phys Lett A 1995;197(5--6):407--12.

\bibitem{SH2a} Torner L, Mihalache D, Mazilu D, Wright EM, Torruellas WE,
Stegeman GI. Stationary trapping of light beams in bulk second-order
nonlinear media. Opt Commun 1995;121(4-6):149--55.

\bibitem{SH2b} Mihalache D, Lederer F, Mazilu D, Crasovan LC.
Multiple-humped bright solitary waves in second-order nonlinear media. Opt
Eng 1996;35(6):1616--23.

\bibitem{SH2c} Malomed BA, Drummond P, He H, Berntson A, Anderson D, Lisak
M. Spatiotemporal solitons in multidimensional optical media with a
quadratic nonlinearity. Phys Rev E 1997;56(4):4725--35.

\bibitem{SH2d} Mihalache D, Mazilu D, Malomed BA, Torner L. Asymmetric
spatio-temporal optical solitons in media with quadratic nonlinearity. Opt
Commun 1998;152(4):365--70.

\bibitem{SH3} Yang JK, Malomed BA, Kaup DJ. Embedded solitons in
second-harmonic-generating systems. Phys Rev Lett 1999;83(10):1958--61.

\bibitem{SHreview0} Etrich C, Lederer F, Malomed BA, Peschel T, Peschel U.
Optical solitons in media with a quadratic nonlinearity. Prog Opt
2000;41:483--568.

\bibitem{SH-review1} Buryak AV, Trapani PD, Skryabin DV, Trillo S. Optical
solitons due to quadratic nonlinearities: from basic physics to futuristic
applications. Phys Rep 2002;370(3):63--235.

\bibitem{SH4} Klein MW, Enkrich C, Wegener M, Linden S. Second-harmonic
generation from magnetic metamaterials. Science 2006;313(5786):502--4.

\bibitem{SH5} Sakaguchi H, Malomed BA. Vortical light bullets in
second-harmonic-generating media supported by a trapping potential. Opt
Express 2013;21(8):9813--23.

\bibitem{SH6} Susanto H, Malomed BA. Embedded solitons in
second-harmonic-generating lattices. Chaos Solitons Fractals
2021;142(7):110534.

\bibitem{Riesz} Agrawal OP. Fractional variational calculus in terms of
Riesz fractional derivatives. J Phys A: Math Theor 2007;40(24):6287.

\bibitem{Riesz2} Cai M, Li CP. On Riesz derivative. FCAA 2019;22(2):287--301.

\bibitem{VK} Vakhitov MG, Kolokolov AA. Stationary solutions of the wave
equation in a medium with nonlinearity saturation. Radiophys Quantum
Electron 1973;16(7):783--9.

\bibitem{Berge} Berg\'{e} L. Wave collapse in physics: principles and
applications to light and plasma waves. Phys Rep 1998;303(5):259--370.

\bibitem{YangJK-Book} Yang JK. Nonlinear Waves in Integrable and
Nonintegrable Systems. SIAM; 2010.
\end{thebibliography}
\end{document}